\documentclass[multphys,vecphys]{svmult}

\usepackage{makeidx}         
\usepackage{graphicx}        
\usepackage{multicol}        
\usepackage[bottom]{footmisc}

\makeindex             

\newfont{\sfsl}{cmssqi8 scaled 1200}

\begin{document}

\title*{Studying the Nature of Dark Energy with Galaxy Clusters}
\author{Thomas H. Reiprich\inst{1}
\and
Daniel S. Hudson\inst{1}
\and
Thomas Erben\inst{1}
\and
Craig L. Sarazin\inst{2}
}
\authorrunning{T.H. Reiprich et al.}
\institute{Argelander-Institut f\"ur Astronomie,\footnote{AIfA was founded by
merging of the Institut f\"ur Astrophysik und Extraterrestrische Forschung, the
Sternwarte, and the Radioastronomisches Institut der Universit\"at Bonn.} Auf
dem H\"ugel 71, 53121 Bonn, Germany
\texttt{thomas@reiprich.net}
\and Department of Astronomy, University of Virginia, P.O. Box 3818,
Charlottesville, VA 22903-0818, USA
}
%
%
\maketitle

\begin{abstract}
We report on the status of our effort to constrain the nature of dark energy
through the evolution of the cluster mass function. {\it Chandra}
temperature profiles for 31 clusters from a local cluster sample are shown. The
X-ray appearance of the proto supermassive binary black hole at the center of
the cluster Abell 400 is described. Preliminary weak lensing results obtained
with Megacam@MMT for a redshift $z=0.5$ cluster from a distant cluster sample
are given.
\end{abstract}

\section{Introduction}
\label{reip:intro}

Understanding the nature of dark energy is one of the major goals of
contemporary cosmological and particle physics. The fate of the universe seems
to be entirely determined by dark energy and a deeper understanding of its
properties may shed light on the unification of general relativity and quantum
theory.

At the moment, astronomical measurements seem most likely to provide further
information about dark energy. Measurements of the evolution of the galaxy
cluster abundance are among the most promising tools; they have the potential 
to yield tight constraints on the equation of state of dark energy. Here, we
report on the status of our project to measure the evolution of the cluster mass
function with very high quality observations of moderately sized local and
distant cluster samples.


\section{Local Cluster Sample}
\label{reip:local}
\begin{figure}
\centering
\includegraphics[height=8cm,angle=270]{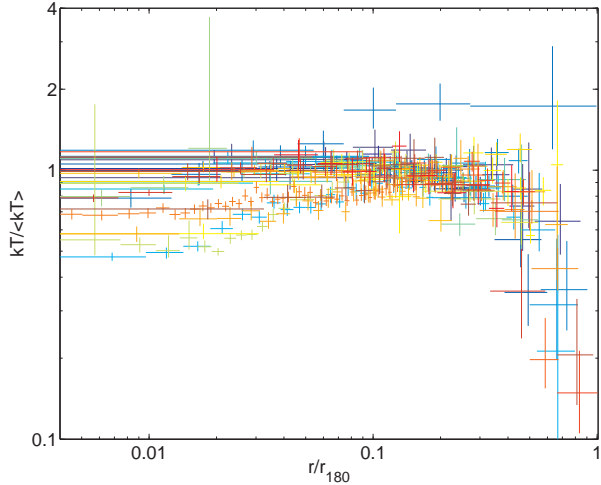}
%
%
\caption{Normalized intracluster gas temperature profiles for 31 clusters
measured with {\it Chandra}.
\vspace{-0.3cm}
}
\label{reip:t-prof}       
\end{figure}
X-ray selection is well suited for the construction of complete cluster samples
useful for cosmological tests, because the X-ray luminosity of clusters is
tightly correlated with their total gravitational mass \cite{reip:rb01,reip:mje05}.
The HIghest X-ray FLUx Galaxy Cluster Sample ({\sfsl HIFLUGCS}) has been selected from the
{\it ROSAT} All-Sky Survey and contains the X-ray brightest galaxy clusters in the sky,
excluding $\pm$20 deg around the Galactic plane as well as the regions of the
Virgo cluster and the Magellanic clouds \cite{reip:rb01}. All clusters have been
reobserved with {\it Chandra} and almost all with {\it XMM-Newton}. We are currently analyzing these
data to determine the most precise local X-ray galaxy cluster mass function.

Currently, cosmological constraints from X-ray galaxy clusters are limited by
systematic effects, especially those affecting the total mass
determination. The high quality {\it Chandra} and {\it XMM-Newton} observations now allow us to
study the gas and temperature structure of clusters in much greater detail
than previously possible.

\begin{figure}
\centering
\includegraphics[height=4.3cm,angle=0]{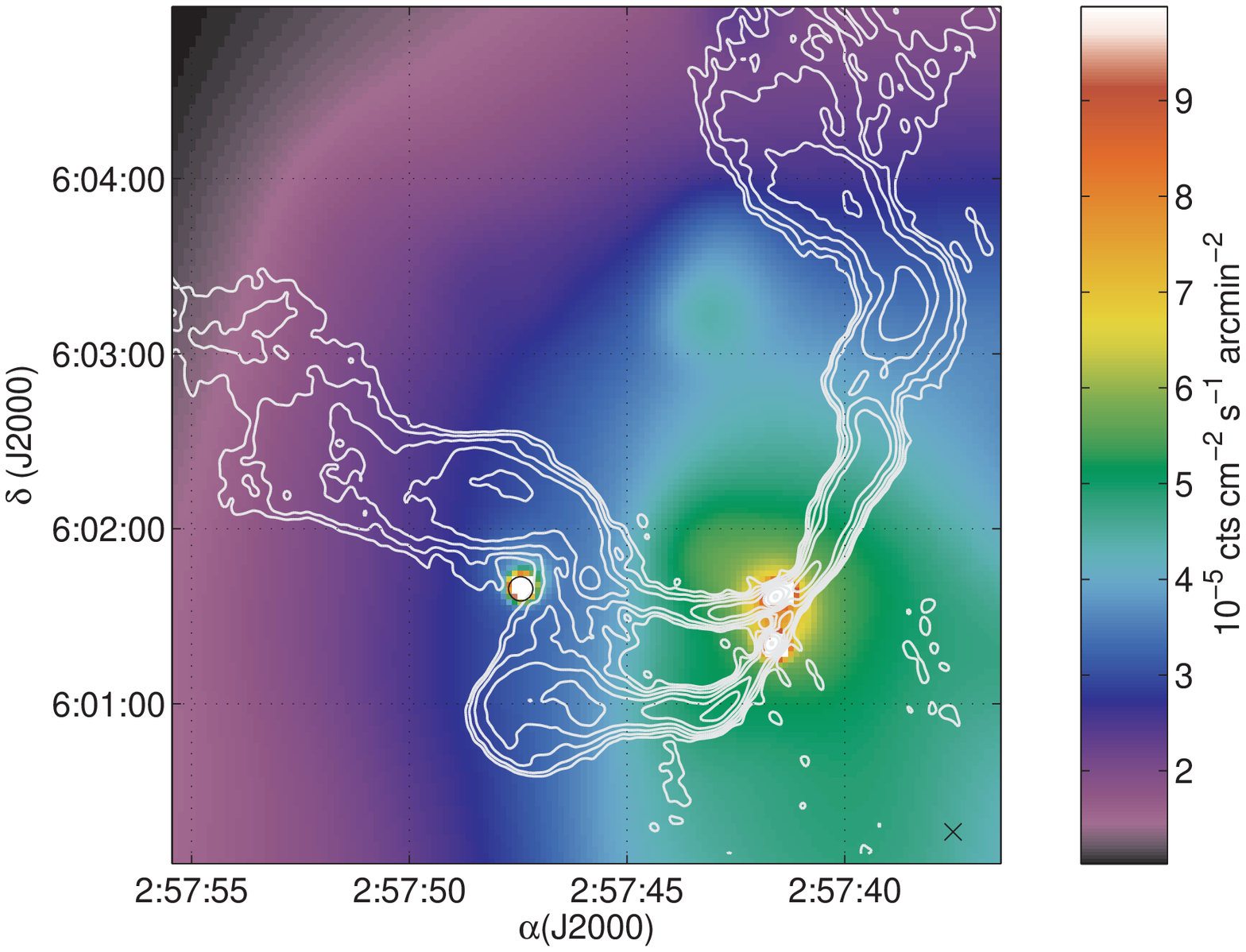}
\includegraphics[height=4.3cm,angle=0]{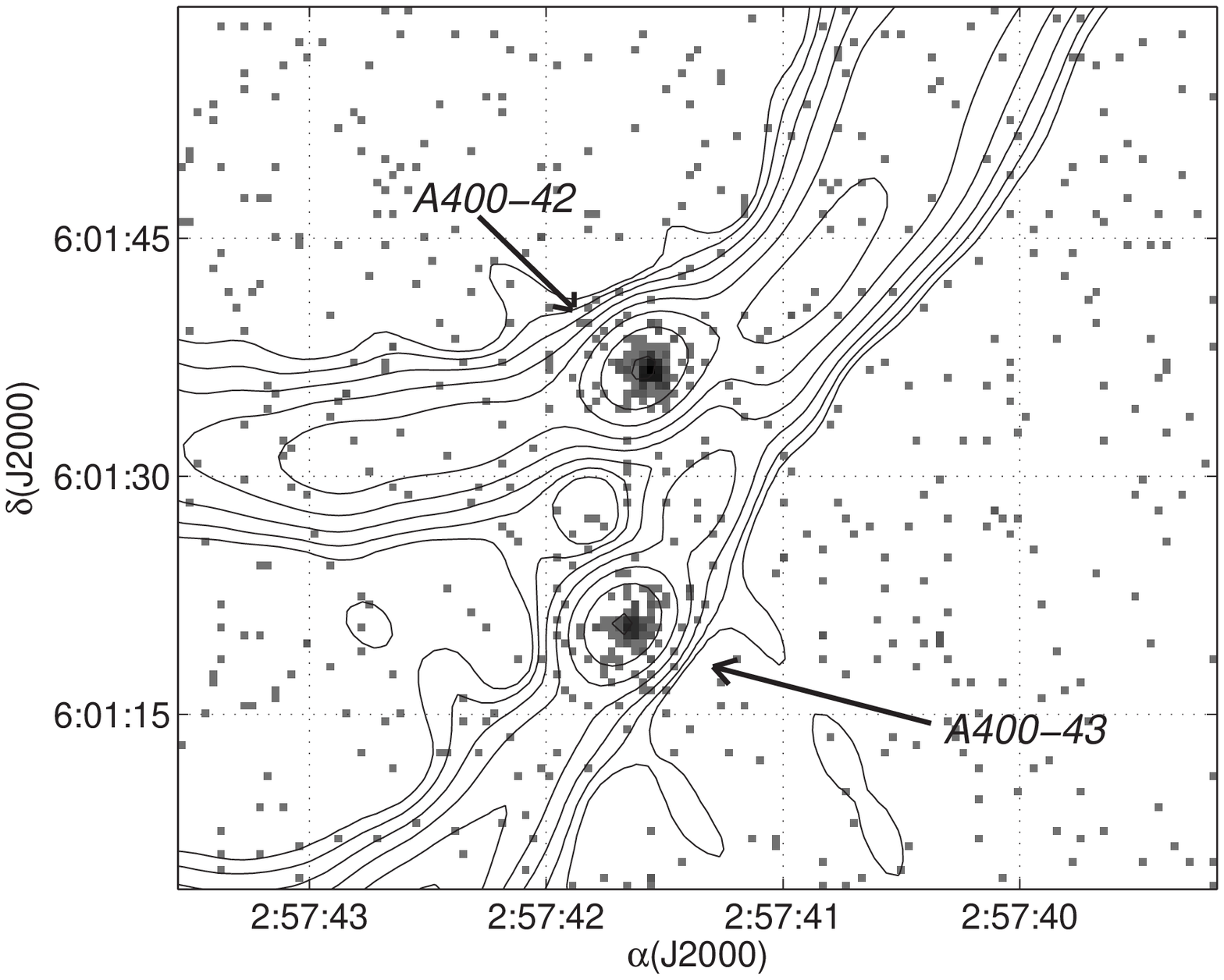}
%
%
\caption{{\it Left:} Adaptively smoothed {\it Chandra} image of the central
region of Abell 400 \cite{reip:hrc06}. Overlaid are the 4.5 GHz VLA radio contours.
{\it Right:} Zooming further into the very center of the unsmoothed {\it
Chandra} image.} 
\label{reip:3C75}       
\end{figure}
Fig.~\ref{reip:t-prof} shows the normalized temperature profiles
of about half of the {\sfsl HIFLUGCS} clusters (Hudson et al., in prep.). A log
scale is used to emphasize details of 
the innermost regions, showing {\it Chandra}'s unrivaled spatial resolution.
This plot gives one of the most detailed views into the statistics of central
temperature structures obtained to date (see also \cite{reip:asf01,reip:vmm05,reip:bjl06}). As
expected, there is no universal temperature profile in the inner part -- some
profiles drop towards the center, others stay flat. Usually, the former are
identified as relaxed clusters and the latter as merging clusters. More
interesting is the fact that even the relaxed clusters do not appear to show a
universal profile in the inner parts -- quite a large spread becomes obvious 
in this log scale plot. This seems contrary to the universal profile for relaxed
clusters that \cite{reip:asf01} found, although they used only 6 clusters. Scatter
was also seen by \cite{reip:vmm05} when they analyzed 13 clusters and, in fact,
considering the complicated physics in cluster centers, non-universality might
be expected. Note that these central regions account for only a small
fraction of the total cluster mass.

The outer radial region where much of the cluster mass resides
($>r_{180}/2$) is rather compressed in this plot. There seems to be a clear
indication for a temperature drop towards larger radii as observed by a number
of previous works but the uncertainties become large, too. We aim to tighten
these constraints by cross correlating the {\it Chandra} data with our {\it
XMM-Newton} results for the same clusters, taking advantage of {\it
XMM-Newton}'s higher throughput and larger field-of-view. 

Analyzing the data for all clusters in this sample, it is no surprise that
exciting details about individual clusters are discovered. As an example, we
show in Figs.~\ref{reip:3C75} the very center of the galaxy cluster A400
\cite{reip:hrc06}, where the well-known radio source 3C75 -- exhibiting a pair of
double radio jets -- resides. Both AGNs, separated by 15 arcsec corresponding to a
projected separation of 7 kpc, are detected separately for the first time in
X-rays. Detailed analysis of the X-ray data reveals further evidence that the 
two AGNs are physically close to each other and form a bound system -- a proto
supermassive binary black hole moving through the intracluster medium
at the supersonic speed of about 1200 km/s.

\section{Distant Cluster Sample}
\label{reip:distant}
The 400 Square Degree (400d) {\it ROSAT} Survey (Burenin et al., in prep.)\ is
the continuation of the 160d Survey \cite{reip:vmf98,reip:mmq03}. It contains clusters
serendipitously detected in basically all useful {\it ROSAT} PSPC pointed
observations. The covered search volume is larger than the volume of the entire
local ($z<0.1$) Universe. A complete subsample of the 41 most luminous, most
distant clusters has been observed with {\it Chandra} in a large program.

\begin{figure}
\centering
\includegraphics[height=5.5cm,angle=0]{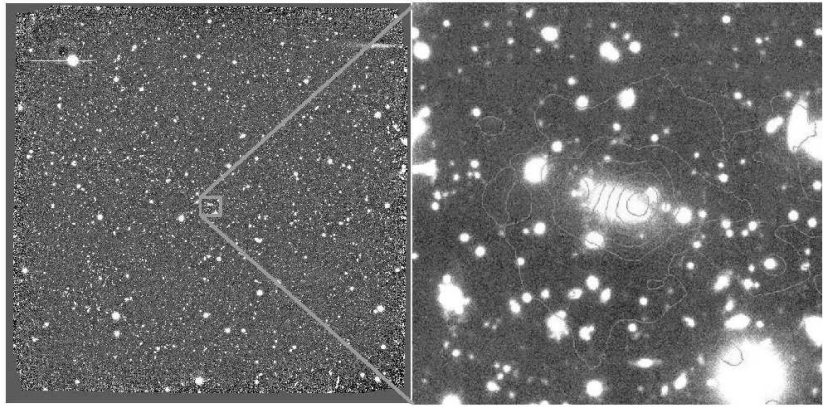}
%
%
\caption{{\it Left:} Coadded Megacam r'-band image of a $z=0.5$ cluster (box
size $\sim$30 arcmin). {\it Right:} Zoom into the very center. {\it Chandra} X-ray
surface brightness contours are overlaid. Note the arc to the right of the
bright central galaxy.}
\label{reip:zoom}       
\end{figure}
\begin{figure}
\centering
\includegraphics[height=8cm,angle=270]{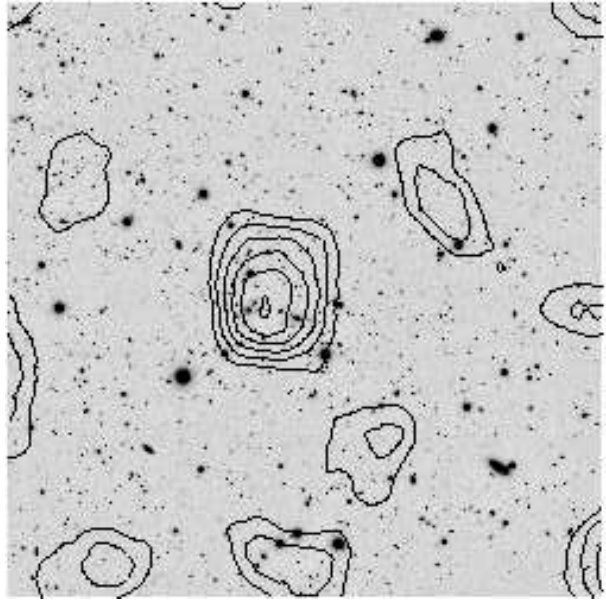}
%
%
\caption{Preliminary weak lensing contours (box size $\sim$10 arcmin). Contours
indicate signal-to-noise ratio of 1.0, 1.5, .., 3.5; the cluster is clearly
detected.}
\label{reip:wl}       
\end{figure}

Since the fraction of galaxy clusters undergoing a major merger is expected to
increase with redshift \cite{reip:cw05}, it would be ideal to have additional
cluster mass estimates -- independent of their dynamical state -- for
these distant clusters. Therefore, we have engaged in a complete weak lensing
follow-up of all 41 clusters. Observations of a redshift $z=0.5$ cluster are
shown as an example in Figs.~\ref{reip:zoom} (Erben et al., in prep.). The
images were taken at 
the 6.5m MMT telescope with the 36-CCD Megacam camera. The images of the 25
dithered observations were reduced and combined with the GaBoDS pipeline
\cite{reip:esd05}, adapted to Megacam. Zooming into the very center, a giant arc
is detected. A preliminary weak lensing analysis shows that the cluster is
clearly detected (Fig.~\ref{reip:wl}), demonstrating for the first time that
Megacam is ideally suited to perform weak lensing measurements for distant
clusters.


\index{HIFLUGCS}
%
{\it Acknowledgements:} This work was supported in part by NASA through {\it
Chandra} Award GO4-5132X and {\it XMM-Newton} Grant
NNG05GO50G as well as the DFG through Emmy Noether Research Grant RE
1462.
The MMT observations were supported in part by the F. H. Levinson Fund
of the Peninsula Community Foundation.
%
%
%



\printindex
\end{document}